\documentstyle[preprint,aps,prd,eqsecnum]{revtex}
\tighten

\def\bfl{\begin{flushleft}}
\def\efl{\end{flushleft}}
\def\bfr{\begin{flushright}}
\def\efr{\end{flushright}}
\def\bc{\begin{center}}
\def\ec{\end{center}}
\def\be{\begin{equation}}
\def\ee{\end{equation}}
\def\ba{\begin{eqnarray}}
\def\ea{\end{eqnarray}}
\def\nn{\nonumber }
\def\lb#1{\label{#1}}

\def\drm{d}
\def\Adequa{\Longleftrightarrow}
\def\schrod{Schr\"odinger  }
\def\vphi{\varphi}
\def\phik#1#2{\varphi_{#1}^{(s)\,#2}}
\def\Mass{{\cal M}}

\def\Der#1#2{\,\frac{\partial #1}{\partial #2}}

\def\Sech#1#2{\, \text{sech}^{#1}\left(#2 \right) }
\def\Cosech#1#2{\, \text{cosech}^{#1}\left(#2 \right) }

\def\HypGF#1#2#3#4{\,\text{F}\left(#1,\,#2,\,#3;\,#4 \right) } 

\def\Cm#1{\, \text{cosh}^{#1}\left( m \rho \right) }
\def\Sm#1{\, \text{sinh}^{#1}\left(m\rho \right) }

\begin{document}

\draft

\title{ 
{\small
~~~~~~~~~~~~~~~~~~~~~~~~~~~~~~~~~~~~~~~~
hep-th/9912063\\
~~~~~~~~~~~~~~~~~~~~~~~~~~~~~~~~~~~~~~~~~~
~~~~~~~~~~~~~~~~~~~~~~~~~~~~~
Phys. Rev. D 61, No. 12, 125017 (2000)}\\ 
Field-to-particle transition based on the
zero-brane approach to quantization of multiscalar field theories
and its application for Jackiw-Teitelboim gravity
      }

\author{Konstantin G. Zloshchastiev\\
~~}
\address{
Department of Physics, National University of Singapore,
Singapore 119260\thanks{On 
leave from Metrostroevskaya 5/453, Dnepropetrovsk 49128, Ukraine}
\\
~~}

\date{~Received:  8 Dec 1999 (LANL), 7 Jan 2000 (PRD) ~}
\maketitle

\begin{abstract}
The field-to-particle transition formalizm based on the effective 
zero-brane action approach
is generalized for arbitrary multiscalar fields.
As a fruitful example, 
by virtue of this method we derive the non-minimal particle action for 
the Jackiw-Teitelboim gravity at fixed gauge in the vicinity of the
black hole solution as an instanton-dilaton doublet.
When quantizing it as the theory with higher derivatives,
it is shown that the appearing quantum equation has SU(2) 
dynamical symmetry group realizing the exact spin-coordinate 
correspondence.
Finally, we calculate the quantum corrections to the 
mass of the JT black hole.
\end{abstract}



\narrowtext

\section{Introduction}\lb{s-i}

The first and foremost
aim of this paper is to develop the classical and 
quantum field-to-particle transition formalism 
for multiscalar field theory in two-dimensional flat spacetime.
Below we will call this formalism as ``zero-brane'' in the sense
``non-minimal point particle'' rather than in the sense 
of supersymmetric string/brane theories.
The study of the field-to-particle transition formalism as such
also lies within the well-known dream programm of constructing of the 
theory which would not contain matter as external postulated entity but
would consider fields as sources of matter and particles as
special field configurations.
Such a programm was inspired probably 
first by Lorentz and Poincar\'e and since that time many efforts had been
made to come it into reality, one may recall the Einstein's, Klein's
and Heisenberg's attempts, 
but discovering of fermionic fields and success of Standard Model
decreased interest to such theories.
Nevertheless, the problem of the origin of matter 
remains to be still open and important, 
especially in what concerns the theoretical explanation of
the fundamental properties of observable particles.
Nowadays, it seems to be possible to realize this programm for 
boson fields (below we will show that 
it is easily possible for multiscalar field in two dimensions) but it
is yet unclear how to obtain fermionic matter.
Some hope that it can be done arises from the supersymmetry but
the generalization of the proposed approach on higher-dimensional
theories encounters severe mathematical troubles.

But thinking about physical relevance of the presented approach
one should not exclude the second possibility. 
Namely, as we will demonstrate below 
the special field solutions indeed can
be correctly regarded as particles which
are sufficiently {\it point-like} ones 
(despite the presence in action of non-minimal terms
depending on curvature, etc.).
From the other hand, there exist the extended models of particles
which suggest that pointness is no more than scale approximation 
\cite{zlo005}.
Thus, one can ask the question of justified choice 
between ``nonminimal point-particle''
and ``extended particle'' paradigmes.
Recently we cannot answer this question yet, we just
point out that this paper is devoted to the former point of view.

Then, by way of an example, we will apply the
field-particle approach to the 
particular 2D theory of gravity admitting
at certain parametrization the correspondence to some scalar field 
theory acting on flat spacetime.
The Jackiw-Teitelboim dilaton gravity discovered in 1984 \cite{jt}
can be obtained also as a dimensional reduction of the 3D BTZ black
hole \cite{btz,ao} and spherically symmetric solution of 4D
dilaton Einstein-Maxwell gravity used as a model of
evaporation process of a near-extremal black hole \cite{cm}.
In spite of the fact taht the 
JT solution is locally diffeomorphic to the
DeSitter space, it has all the global attributes of a black hole.
Besides, it is simple enough to obtain main results in a 
non-perturbative way that seems to be important for highly non-linear 
general relativity. 

Despite the wide literature devoted to classical and quantum
aspects of the theory 
(see Refs. \cite{gk,lgk,dil-qua} and references therein),
it is concerned mainly with standard methods of study, whereas it
is clear that black holes are extended objects and thus should
be correctly considered within framework of the brane theory
where there is no rigid fixation of spatial symmetry.
The quantum aspects were studied mainly in connection with
group features of JT dilaton gravity in general 
whereas we will quantize the theory 
in the vicinity of certain non-trivial vacuum induced by a static solution 
emphasizing on the corrections to mass spectrum.
Thus, our purpose is to study the 2D dilaton gravity
in the neighbourhood of the 
classical and quantum Jackiw-Teitelboim black hole 
within the frameworks of the brane approach, which 
consists in the 
constructing of the effective action where the non-minimal
terms (first of all, depending on the world-volume curvature) are induced
by field fluctuations.
Then the effective action evidently arises after nonlinear 
reparametrization
of an initial theory when excluding zero field oscillations.

The paper is arranged as follows.
In Sec. \ref{s-jt} we study the JT solution as a soliton-dilaton
(more correctly, instanton-dilaton) doublet and 
its properties at the classical level.
In Sec. \ref{s-ea} we generalize the approach \cite{kpp} for 
arbitrary multiscalar fields and apply it for JT dilaton gravity in 
fixed-gauge (flat-spacetime) formulation.
Minimizing the action with respect to field fluctuations,
we remove zero modes and obtain the point-particle action
with non-minimal terms corresponding to this theory. 
Sec. \ref{s-q} is devoted to quantization of the action as a constrained
theory with higher derivatives.
In result we obtain the \schrod wave equation describing
wave function and mass spectrum of a point particle with curvature
and apply them for the JT black hole.
Then we calculate the zeroth and first excited levels to get the
mass of the quantum JT black hole with quantum corrections.
Conclusions are made in Sec. \ref{s-c}.

\section{Jackiw-Teitelboim gravity}\lb{s-jt}

Consider the action of Jackiw-Teitelboim dilaton gravity 
\begin{equation}
S_{JT}[\tau, g] = \frac{1}{2 G} \int \drm^2 x \sqrt{-g} 
\tau (R+2 m^2),                                               \lb{eq2.1} 
\end{equation}
where $G$ is the gravitational coupling constant, dimensionless in 
2D case. Extremizing this action with respect to metric and dilaton
field variations we obtain the following equations of motion
\ba                                                           
&&R + 2 m^2= 0,                                                \lb{eq2.2}\\
&&\left(\nabla_\mu\nabla_\nu - m^2 g_{\mu\nu} \right)\tau= 0.  \lb{eq2.3}
\ea
Further, if one performs the parametrization of a metric
\be                                                            \lb{eq2.4}
\drm s^2 = - \sin{\!^2 (u/2)} \drm t^2 + 
\cos{\!^2 (u/2)} \drm x^2,
\ee
and puts the metric ansatz into
eqs. (\ref{eq2.1})-(\ref{eq2.3}), they can be rewritten \cite{gk}, 
respectively, as
\begin{equation}
S_{JT}[\tau, u] = \frac{1}{2 G} \int \drm^2 x 
\tau (\Delta u - m^2 \sin{u}),                                \lb{eq2.5} 
\end{equation}
\ba                                                           
&&\Delta u - m^2 \sin{u}= 0,                         \lb{eq2.6}  \\
&&\left(\Delta - m^2 \cos{u} \right)\tau= 0,  
\ea
where $\Delta$ is the flat Euclidean Laplacian, 
$\partial_{t}^2 + \partial_{x}^2$.

If we wish to choose from the solutions of eq. (\ref{eq2.6}) only the 
one-instanton ones, we have the following 
instanton-dilaton pair:
\ba                                                           
&&u^{(s)} (x,t) = 4 \arctan{\exp{(m\rho)}},             \lb{eq2.8}  \\
&&\tau^{(s)} (x,t) = -C_1 \Sech{}{m\rho} +
C_2 
\left[
\sinh{(m\rho)} + m\rho \Sech{}{m\rho}
\right],                                                \lb{eq2.9}  
\ea
where $C_i$ are arbitrary constants, $\rho = \gamma (x-v t)$, 
$1/\gamma = \sqrt{1 + v^2}$. Then the metric (\ref{eq2.4}) after
the transformation $\{x,t\} \to \{R,T\}$, such that
\ba
&&\drm T = v \sqrt{\frac{m}{2 G \gamma M}}
\left[
      \drm t - \frac{v/\gamma}{\Cosech{2}{m\rho}- v^2}\drm \rho
\right],   \nn\\
&&R = \frac{1}{v} \sqrt{\frac{2 G M}{\gamma m^3}}
\Sech{}{m\rho}, \nn
\ea
where $M$ is an arbitrary constant, can be rewritten in the explicit
form representing the JT black hole solution
\be                                                            \lb{eq2.10}
\drm s^2 = 
- 
\left(
       m^2 R^2 - \frac{2 G \gamma M}{m}
\right)\drm T^2 
+ 
\left(
       m^2 R^2 - \frac{2 G \gamma M}{m}
\right)^{-1} \drm R^2, 
\ee
having the following energy and event horizon
\be                                                        \lb{eq2.11}
E_{\text{BH}} = \gamma M, \ \
R_{\text{BH}} = \sqrt{\frac{2 G \gamma M}{m^3}}.
\ee
Our purpose now will be to take into account the field fluctuations
in the neighborhood of this solution and to construct the effective
action of the JT black hole as a zero-brane.
Before we go further, one should develop a general approach.

\section{Effective action}\lb{s-ea}

In this section we will construct the nonlinear effective action of 
an arbitrary multiscalar 2D theory in the vicinity of 
a solitary-wave solution, and then apply it for JT dilaton gravity.
In fact, here we will
describe the procedure of the correct transition from field 
to particle degrees of freedom.
Indeed, despite the solitary-wave solution resembles a particle
both at classical and quantum levels,
it yet remains to be a {\it field solution}  with infinite number of
field degrees of freedom
whereas a true particle has a finite number of degrees of freedom.
Therefore, we are obliged to correctly handle this circumstance
(and several others)
otherwise deep contradictions may appear.

\subsection{General formalism}\lb{s-ea-gf}

Let us consider the action describing $N$ scalar fields
\begin{equation}
S[\vphi] = \int  L(\vphi)\, \drm^2 x,                        \lb{eq3.1} 
\end{equation}
\be                                                          \lb{eq3.2}
L (\vphi) = \frac{1}{2} 
\sum_{a=1}^{N} (\partial_m \varphi_a) (\partial^m \varphi_a) -
U (\vphi).
\ee
The corresponding equations of motion are
\be                                                          \lb{eq3.3}
\partial^m \partial_m \varphi_a + U_a(\vphi) = 0,
\ee
where we defined                                   
\[
U_a(\vphi) = \Der{U(\vphi)}{\varphi_a},~~
U_{ab}(\vphi) = \Der{^2\, U(\vphi)}{\varphi_a \partial \varphi_b}.
\]
Suppose, we have a solution in the class of solitary waves
\be                                                           \lb{eq3.4}
\phik{a}{}(\rho) = \phik{a}{}
\left( \gamma ( x-v t)
\right),
~~\gamma= 1/\sqrt{1-v^2},
\ee
having the localized energy density
\be                                                           \lb{eq3.5}
\varepsilon (\vphi) = \sum_{a} 
\Der{L(\vphi)}{(\partial_0\vphi_a)} \partial_0\vphi_a - L(\vphi),
\ee
and finite mass integral
\be                                                            \lb{eq3.6}
\mu = \int\limits_{-\infty}^{+\infty}
\varepsilon (\phik{}{})\ \drm \rho =
-\int\limits_{-\infty}^{+\infty}
L (\phik{}{})\ \drm \rho < \infty,
\ee
coinciding with the total energy up to the Lorentz factor $\gamma$.

Let us change to the set of the collective coordinates 
$\{\sigma_0=s,\ \sigma_1=\rho\}$ such that
\be                                                       \lb{eq3.7}  
x^m = x^m(s) + e^m_{(1)}(s) \rho,\ \              
\varphi_a(x,t) = \widetilde \varphi_a (\sigma),     
\ee
where $x^m(s)$ turn out to be the coordinates of a (1+1)-dimensional point
particle, $e^m_{(1)}(s)$ is the unit spacelike vector orthogonal
to the world line.
Hence, the action (\ref{eq3.1}) can be rewritten in new coordinates as
\be                                                         \lb{eq3.8}
S[\widetilde \varphi] = 
\int L (\widetilde \varphi) \,\Delta \ \drm^2 \sigma,
\ee
\[
L (\widetilde \varphi) = \frac{1}{2} \sum_{a}
\left[
      \frac{(\partial_s \widetilde\varphi_a)^2}{\Delta^2} - 
                  (\partial_\rho \widetilde\varphi_a)^2
\right]
- U (\widetilde \varphi),
\]
where
\[
\Delta = \text{det} 
\left|
\left|
      \Der{x^m}{\sigma^k}
\right|
\right|
= \sqrt{\dot x^2} (1- \rho k),
\]
and $k$ is the curvature of a particle world line
\be                                                            \lb{eq3.9}
k = \frac{\varepsilon_{mn} \dot x^m \ddot x^n}{(\sqrt{\dot x^2})^3},
\ee
where $\varepsilon_{m n}$ is the unit antisymmetric tensor.
This new action contains the $N$ redundant degrees of freedom which 
eventually
lead to appearance of the so-called ``zero modes''.
To eliminate them we must constrain the model
by means of the condition of vanishing of the functional derivative with
respect to field fluctuations about a chosen static solution,
and in result we will obtain the required effective action.

So, the fluctuations of the fields $\widetilde\varphi_a (\sigma)$ in the 
neighborhood of the static solution $\phik{a}{} (\rho)$
are given by the expression
\be                                               
\widetilde\varphi_a (\sigma) = 
\phik{a}{} (\rho) + \delta \varphi_a (\sigma).
\ee
Substituting them into eq. (\ref{eq3.8}) and considering the static
equations of motion (\ref{eq3.3}) for $\phik{a}{}(\rho)$ we have
\ba                                          \lb{eq3.11}
S[\delta \vphi] 
&=& \int d^2 \sigma \ 
   \Biggl\{\Delta 
        \Biggl[L( \phik{}{}) +
               \frac{1}{2} \sum_{a}
               \Biggl( 
                  \frac{\left(\partial_s \ \delta \varphi_a \right)^2}
                       {\Delta^2} 
                  -  
                  \Bigl( 
                        \partial_{\rho}  \delta \varphi_a 
                  \Bigr)^2 - \nn\\
&&                  \sum_b U_{ab} ( \phik{}{}) 
                  \delta \varphi_a\, \delta \varphi_b
               \Biggr) 
        \Biggr]
        - k \sqrt{\dot x^2}\sum_a \phik{a}{\prime} \delta \varphi_a
        + O (\delta \varphi^3)                            
    \Biggr\} + \{\text{surf. terms}\},                                            
\ea
\[
L( \phik{}{}) = -
 \frac{1}{2} \sum_a \phik{a}{\prime\, 2} - U ( \phik{}{}),
\]
where prime means the derivative with respect to $\rho$.
Extremizing this action with respect to 
$\delta \varphi_a$ one can obtain 
the system of equations in partial derivatives for field fluctuations:
\be
\left(
     \partial_s \Delta^{-1} \partial_s -
     \partial_{\rho} \Delta \partial_{\rho} 
\right) \delta\varphi_a
+\Delta \sum_{b} U_{ab} ( \phik{}{}) \delta\varphi_b
+ \phik{a}{\prime} k\sqrt{\dot{x}^2} =
O(\delta \varphi^2),
\ee
which has to be the constraint removing redundant degrees of
freedom.
Supposing $\delta\varphi_a (s,\rho) = k(s) f_a(\rho)$, in the 
linear approximations
$\rho k\ll 1$ (which naturally guarantees also
the smoothness of a world line at $\rho \to 0$) 
and $O(\delta\varphi^2)=0$ we obtain the system
of $N+1$ ordinary derivative equations
\ba  
&&\frac{1}{\sqrt{\dot{x}^2}} \frac{d}{ds} 
\frac{1}{\sqrt{\dot{x}^2}} \frac{dk}{ds} +ck = 0,          \lb{eq3.13}\\
&&-f''_a + \sum_b
\left( 
      U_{ab} ( \phik{}{}) - c \delta_{ab}
\right) f_b + \phik{a}{\prime} = 0,                        \lb{eq3.14}
\ea
where $c$ is the constant of separation.
Searching for a solution of the last subsystem in the form
\be                                                       \lb{eq3.15}
f_a = g_a + \frac{1}{c} \phik{}{\prime},
\ee
we obtain the homogeneous system
\be  
-g''_a + \sum_b
\left( 
      U_{ab} ( \phik{}{}) - c \delta_{ab}
\right) g_b = 0.                                           \lb{eq3.16}
\ee
Strictly speaking, the explicit form of $g_a (\rho)$ is not significant 
for us, because we always can suppose integration constants to be zero
thus restricting ourselves by the special solution.
Nevertheless, the homogeneous system should be considered as the 
eigenvalue problem for $c$ (see below).

Substituting the found functions $\delta\varphi_a = k f_a$ back 
in the action 
(\ref{eq3.11}), we can rewrite it in the explicit zero-brane form
\be                                \lb{eq3.17}
S_{\text{eff}} = 
S_{\text{eff}}^{\text{(class)}} + S_{\text{eff}}^{\text{(fluct)}} =
- \int \drm s \sqrt{\dot x^2} 
\left(
       \mu + \alpha k^2
\right),
\ee
describing a point particle with curvature,
where $\mu$ was defined in (\ref{eq3.6}), and
\be                                               \lb{eq3.18}
\alpha = 
\frac{1}{2} \sum_a \int\limits_{-\infty}^{\infty} 
f_a \phik{a}{\prime} \ \drm \rho
+
\frac{1}{2} \sum_a 
\int\limits_{-\infty}^{+\infty} 
\left(
f_a f_a^{\prime}
\right)^\prime \ \drm \rho.
\ee
Further, contracting (\ref{eq3.3}) with $\phik{a}{\prime}$, we
obtain the expression
\be                                                \lb{eq3.19}
\sum_a 
\left(
\phik{a}{\prime\prime} -
U_a(\phik{}{})
\right) 
\phik{a}{\prime} = 0,
\ee
which can be rewritten as 
\be                                                \lb{eq3.20}
\sum_a 
\phik{a}{\prime 2} 
= 2 U(\phik{}{}(\rho)).
\ee
Considering the Eqs. (\ref{eq3.5}), (\ref{eq3.6}), (\ref{eq3.15}), 
(\ref{eq3.19}) and (\ref{eq3.20}), the expression for $\alpha$ can be 
written in the simple form (for simplicity here we suppose  the
same eigenvalues, $c_a \equiv c$, otherwise 
the first integral in Eq. (\ref{eq3.18})) cannot be reduced 
to the integral (\ref{eq3.6})
and should be evaluated separately)
\be
\alpha = \frac{\mu}{2 c} + \frac{1}{2 c^2}
\int\limits_{-\infty}^{+\infty} 
U^{\prime\prime}(\phik{}{}(\rho)) \ \drm \rho,
\ee
where the second term can be integrated as a full derivative.
Therefore, even if it is non-zero\footnote{It 
identically vanishes when $|\phik{a}{}(\rho)| \leq O(1)$ at infinity.},
we always can remove it by means of including into surface terms of
the action (\ref{eq3.11}) or adding of an appropriate
counterterm to the action (\ref{eq3.8}):
\[
S^{\text{(reg)}} [\widetilde \varphi]  =
S [\widetilde \varphi] - 
\frac{1}{2 c^2}
\int\limits_{-\infty}^{+\infty}\, \drm^2 \sigma \Delta k^2
U^{\prime\prime}(\rho).
\] 
Thus, we obtain the final form of the effective zero-brane action of
the theory
\be                                \lb{eq3.22}
S_{\text{eff}} = 
- \mu \int \drm s \sqrt{\dot x^2} 
\left(
       1 + \frac{1}{2 c} k^2
\right).
\ee
It is straightforward to derive the corresponding 
equation of motion in the Frenet basis
\be
\frac{1}{\sqrt{\dot x^2}}
\frac{\drm}{\drm s}
\frac{1}{\sqrt{\dot x^2}}
\frac{\drm k}{\drm s} +
\left(c - \frac{1}{2} k^2
\right) k = 0,
\ee
hence one can see that eq. (\ref{eq3.13}) was nothing
but this equation in the linear approximation $k \ll 1$,
as was expected.

Thus, the only problem which yet demands  
resolving is the determination of eigenvalue $c$.
It turns out to be the Sturm-Liouville problem for the system 
(\ref{eq3.16}) under some chosen boundary conditions.
If one supposes, for instance, the finiteness of $g$ at infinity
then the $c$ spectrum turns out to be discrete.
Moreover, it often happens that $c$ has only one or two admissible
values\footnote{For instance, in the works \cite{kpp} 
(one-scalar $\varphi^4$ theory) or \cite{zlo006} 
($\varphi^3$ and Liouville model),
where the special cases of this formalism were used, $c$ has the form
$\beta m^2$ where $\beta$ is a single positive half-integer or integer;
the cases with $c<0$ does not have, as a rule, independent physical
sense, because at quantization they either can be interpreted in terms
of antiparticles or appear to be unphysical at all.}.

Be that as it may, the exact value of $c$ is necessary hence the system 
(\ref{eq3.16}) should be resolved as exactly as possible.
Let us consider it more closely.
The main problem there is the functions $g_a$ are mixed between
equations.
To separate them, let us recall that there exist $N-1$ orbit
equations, whose varying resolves the separation problem.
We consider this for the case $N=2$, i.e., for a biscalar theory,
all the more so it will be helpful when applying for JT gravity.

Considering (\ref{eq3.15}), the varying of a single 
orbit equation yields
\be
\frac{\delta \varphi_2}{\delta\varphi_1} = 
\frac{\phik{2}{\prime}}{\phik{1}{\prime}} =
\frac{g_2}{g_1},
\ee
hence the system (\ref{eq3.16}), $N=2$, can be separated into the
two independent equations
\be  
-g''_a + 
\left( 
      \frac{\phik{a}{\prime\prime\prime}}
           {\phik{a}{\prime}}
      - c 
\right) g_a = 0,                                           \lb{eq3.25}
\ee
if one uses 
$\displaystyle{\varphi_a''' = \sum_b U_{ab} (\varphi) \varphi_b'}$.
In this form it is much easier to resolve the eigenvalue problem.
Therefore, the two independent parameters for the action (\ref{eq3.22}), 
$\mu$ and $c$, can be determined immediately by virtue of
eqs. (\ref{eq3.6}) and (\ref{eq3.25}).

Finally, it should be pointed out that the developed method can be 
generalized both in the qualitative direction (considering it for 
the Yang-Mills and spinor Lagrangians \cite{zlop}) and toward the 
increasing of spatial dimensions.


\subsection{Application for JT black hole}\lb{s-ea-af}

For further studies 
it is convenient to perform the Wick rotation and to
work in terms of solitons and Lorentzian time rather then in terms
of instantons and Euclidean time, all the more so the main results
of the previous subsection are independent of $v$.
Omitting topological surface terms, we will consider instead of the 
action (\ref{eq2.5}) its Lorentzian analogue
\begin{equation}
S_{JT}[\tau, u] = \frac{1}{2 G} \int \drm^2 x 
(\partial_m \tau \partial^m u - 
m^2 \tau \sin{u}).                                          \lb{eq3.26} 
\end{equation}
The soliton-dilaton doublet (\ref{eq2.8}), (\ref{eq2.9}) 
has the localized energy density (\ref{eq3.5})
\[                                                          
\varepsilon (x,t) = \frac{2 m^2}{G}
\frac{
      C_1 \tanh{(m \rho)} +
      C_2
      \left[
            1 - m\rho \tanh{(m \rho)}
      \right]
      }
      {\cosh{\!^2\, (m\rho)}},                                                        
\]
and can be interpreted as the relativistic point particle with the
energy
\be                                                          \lb{eq3.27}
E_{\text{class}} = \int\limits_{-\infty}^{+\infty}
\varepsilon (x,t)\ \drm x \equiv \gamma \mu =
\frac{2 C_2 \gamma m}{G}, 
\ee
i.e. the integral (\ref{eq3.6}) is finite and coincides with the 
energy (\ref{eq2.11})
\be                                                          \lb{eq3.28}
\mu = M,
\ee
if one redefines $C_2$.

The action (\ref{eq3.26}) always can be linearly rearranged in the form 
(\ref{eq3.1}), (\ref{eq3.2}), 
if we introduce fields $\varphi_a$ such that
\be                                                         \lb{eq3.29}
2\theta u = \varphi_1 - i \varphi_2,\quad
\tau/\theta = \varphi_1 + i \varphi_2,
\ee
where $\theta$ is an arbitrary real constant 
which (similarly to $C_1$) will not affect on final results.
We will suppose the final zero-brane
action (\ref{eq3.22}).
The flat spacetime coordinates $x^\mu$ (\ref{eq3.7}) should be understood
in the sense we substituted initial curved spacetime for flat one
with effective potential,
but the meaning of the collective coordinates $\rho$ and $s$
remains unchanged because it describes internal structure and hence
is independent of whether we are
working in curved or flat space.

Therefore, the main task now is to
specify the parameters of the action (\ref{eq3.22})
for our case.
We  have $\mu$ already determined by (\ref{eq3.28}), and 
the eigenvalue $c$ remains to be 
the only unknown parameter for eq. (\ref{eq3.22}).
For eqs. (\ref{eq3.25}) we will require 
the traditional boundary conditions
\be                                                          
g_a(+\infty) - g_a(-\infty) = O (1),
\ee
whereas provided eqs. (\ref{eq2.8}), (\ref{eq2.9}), (\ref{eq3.29}) 
the system (\ref{eq3.25}) have the form
\be  
-g''_a + 
\left( 
      K_a
      - c 
\right) g_a = 0,                                           \lb{eq3.31}
\ee
where
\ba
&&K_1 = \frac{m^2}{\Cm{2}}
\frac{
       A_1 (\Cm{2}-2) + C_2 \Cm{5} + A_2 (6-\Cm{2})
     }
     {
      \Cm{} (4\theta^2+ C_2 + C_2 \Cm{2}) - A_2
     }, \nn\\
&& K_2 = K_1 |_{C_i \to - C_i}, \nn\\
&&A_1 = \Cm{} (4\theta^2 + 3 C_2),   \quad
A_2 = \Sm{} (C_1 + C_2 m \rho), \nn
\ea
hence it is clear that
\be                                               \lb{eq3.32}
K_a (0) = -m^2, \quad 
K_a (-\infty) = K_a (+\infty) = m^2.  
\ee

This eigenvalue equation is evidently hard to solve exactly,
hence we use the method of the approximing potential which would have
the main properties of $K_a$ especially those presented by (\ref{eq3.32}). 
Besides, we will consider the equation for $K_1$ only because both
potentials have approximately the same behaviour.

Thus, omitting the index we will assume the following eigenvalue equation
\be                                                          
-g'' + m^2
\left(
1- \frac{2}{\cosh{\!^2\, (m\rho)}}
\right)g - c g = 0,
\ee
instead of eq. (\ref{eq3.31}).
Its potential has the main features of $K_a$ but appears to be 
exactly solvable:
according to the proven theorem (see Appendix \ref{a-a}), 
the only admissible non-zero $c$ is
\be                                          \lb{eq3.34}
c=m^2.
\ee
This result is confirmed also by the quasiclassical approximation.
Indeed, the necessary condition 
of convergence of the phase integral 
\[
\oint p \drm \rho = 
2 \int\limits_{-\infty}^{+\infty} \sqrt{K_a - c} ~ \drm \rho
\]
appears to be the following one
\[
K_a (\pm \infty) - c =0,
\]
which yields eq. (\ref{eq3.34}) again.

Therefore, the effective zero-brane action of the dilaton gravity about
Jackiw-Teitelboim black hole with fluctuational corrections is
\be                                                   \lb{eq3.35}
S_{\text{eff}} = 
- \mu \int \drm s \sqrt{\dot x^2} 
\left(
       1 + \frac{k^2}{2 m^2} 
\right),\ \ \mu=M=\frac{2 C_2 m}{G}.
\ee
In the next section we will quantize it to obtain the quantum 
corrections to the mass of fixed-gauge JT black hole.

\section{Quantization}\lb{s-q}

In the previous section we obtained a classical 
effective action for the model in question.
Thus, to quantize it we must consecutively construct the 
Hamiltonian structure of dynamics of the point particle with 
curvature \cite{ner,ply,dhot}.

\subsection{General formalism}\lb{s-q-gf}

From the brane action (\ref{eq3.22}) 
and definition of the world-line curvature 
one can see that we have the 
theory with higher derivatives \cite{ply,dhot}.
Hence, below we will treat the coordinates and momenta as the 
canonically independent coordinates of phase space.
Besides, the Hessian matrix constructed 
from the derivatives with respect to accelerations,
\[
M_{a b} = 
\left|
\left|
\Der{^2 L_{\text{eff}}}{\ddot x^a \partial \ddot x^b}
\right|
\right|,
\] 
appears to be singular that says about the presence of the
constraints on phase variables of the theory.

As was mentioned, the phase space consists of the two pairs of 
canonical variables:
\ba
&&x_m,\ \ p_m = \Der{L_{\text{eff}}}{q^m} - \dot \Pi_m, \\
&&q_m = \dot x_m,\ \ \Pi_m =\Der{L_{\text{eff}}}{\dot q^m},
\ea
hence we have
\ba
&&p^n = - e^n_{(0)} \mu 
\left[
      1-  \frac{1}{2 c}
\right] +
\frac{\mu}{c}
\frac{e^n_{(1)} }{\sqrt{q^2}} \dot k,  \\
&&
\Pi^n = - 
\frac{\mu}{c}
\frac{e^n_{(1)}}{\sqrt{q^2}} k,
\ea
where the components of the Frenet basis are
\[
e^m_{(0)} = \frac{\dot x^m}{\sqrt{\dot x^2}},\
e^m_{(1)} = - \frac{1}{\sqrt{\dot x^2}} \frac{\dot e^m_{(0)}}{k}.
\]
There exist the two primary constraints of first kind
\ba
&&\Phi_1 = \Pi^m q_m \approx 0, \\
&&\Phi_2 = p^m q_m + \sqrt{q^2} 
\left[
      \mu + \frac{c}{2 \mu} q^2 \Pi^2
\right]  \approx 0,
\ea
besides we should add the proper time gauge condition,
\be
G = \sqrt{q^2} - 1 \approx 0,
\ee
to remove the non-physical gauge degree of freedom.
Then, when introducing the new variables,
\be
\rho = \sqrt{q^2},\ \ v = 
\text{arctanh} 
\left(
      p_{(1)}/p_{(0)}
\right),
\ee
the constraints can be rewritten in the form
\ba
&&\Phi_1 = \rho \Pi_\rho, \nn\\
&&\Phi_2 = \rho 
\left[
      -\sqrt{p^2} \cosh{v} + \mu -
      \frac{c}{2 \mu}
      \left( 
            \Pi^2_v - \rho^2 \Pi^2_\rho
      \right)
\right],                                       \\
&&G=\rho-1, \nn
\ea
hence finally we obtain the constraint
\be
\Phi_2 = 
      -\sqrt{p^2} \cosh{v} + \mu -
      \frac{c}{2 \mu}
      \Pi^2_v \approx 0,                                       
\ee
which in the quantum theory ($\Pi_v = - i \partial/\partial v$) 
yields 
\[
\widehat\Phi_2 |\Psi\rangle =0.
\]
As was shown in Ref. \cite{kpp} (see also Ref. \cite{ply}), 
the constraint $\Phi_2$ on the 
quantum level admits several coordinate representations that,
generally speaking, lead to different 
nonequivalent theories, therefore,
the choice between the different forms of 
$\widehat\Phi_2$ should be based on the physical relevance.
Then the physically admissible
equation determining quantum dynamics of the quantum
kink and bell particles has the form:
\be                                                            \lb{eq4.11}
[ \widehat H-\varepsilon] \Psi(\zeta) = 0, 
\ee 
\be
\widehat H =  -\frac{\drm^2}{\drm \zeta^2} +
  \frac{B^2}{4}
  \sinh{\! ^2 \zeta}
  -B
  \left(
        S+\frac{1}{2}
  \right)
  \cosh{\zeta},                                             
\ee
where
\ba
&&\zeta=v/2,\ \sqrt{p^2} = \Mass, \nn\\
&&B= 8 \sqrt{
            \frac{\mu \Mass}{c}  
            },                                              \lb{eq4.13}\\
&& \varepsilon = \frac{8 \mu^2}{c}  
\left(
      1 - \frac{\Mass}{\mu}
\right), \nn
\ea
and $S=0$ in our case.

As was established in the works \cite{raz,zu}, 
SU(2) has to be the dynamical symmetry
group for this Hamiltonian which can be rewritten in the form of
the spin Hamiltonian
\be
\widehat H= -S^2_z - B S_x,                                 
\ee
where the spin operators,
\ba
&&S_x = S \cosh{\zeta} - \frac{B}{2} \sinh{\!^2 \zeta} - \sinh{\zeta} 
\frac{\drm}{\drm\zeta},  \nn \\
&&S_y = i         \left\{
               -S \sinh{\zeta} + \frac{B}{2} \sinh{\zeta}\cosh{\zeta} + 
\cosh{\zeta} \frac{\drm}{\drm\zeta} \right\},   \\
&&S_z =          \frac{B}{2} \sinh{\zeta}
        + \frac{\drm}{\drm\zeta},              \nn
\ea
satisfy with the commutation relations
\[
[S_i,~S_j] = i \epsilon_{ijk} S_k,                       
\]
besides
\[
S_x^2+S_y^2+S_z^2 \equiv S (S+1).                         
\]
In this connection it should be noted that though the reformulation of 
some interaction 
concerning the coordinate degrees of freedom in terms of
spin variables is widely used (e.g., in the theories with the
Heisenberg Hamiltonian, see Ref. \cite{lp}), it has to be just
the physical approximation as a rule,
whereas in our case the spin-coordinate correspondence is exact.

Further, 
at $S\geq 0$ there exists an irreducible ($2 S+1$)-dimensional 
subspace of the representation space of the su(2) Lie algebra, which is
invariant with respect to these operators.
Determining eigenvalues and eigenvectors of the spin Hamiltonian
in the matrix representation 
which is realized in this subspace, one can prove 
that the solution of eq. (\ref{eq4.11}) is the function
\ba
\Psi (\zeta) =\exp{
                   \left(
                         -\frac{B}{2} \cosh{\zeta}
                   \right)
                  }
              \sum_{\sigma=-S}^{S}
              \frac{c_\sigma}
                   {
                    \sqrt{
                          (S-\sigma)\verb|!|~
                          (S+\sigma)\verb|!|
                         }
                   }
              \exp{
                   \left(
                         \sigma \zeta
                   \right)
                  }, 
\ea
where the coefficients $c_\sigma$ are the solutions of 
the system of linear equations
\[
\biggl(
       \varepsilon+\sigma^2
\biggr)c_\sigma + \frac{B}{2}
\biggl[
       \sqrt{(S-\sigma)(S+\sigma+1)}~ c_{\sigma+1}             
+ \sqrt{(S+\sigma)(S-
\sigma+1)}~ c_{\sigma-1}
\biggr] = 0,
\]
\[
c_{S+1} = c_{-S-1}=0,~~\sigma=-S,~-S+1,...,~S.            
\]
However, it should be noted that these expressions give only the 
finite number of exact solutions which is equal to the dimensionality of
the invariant subspace 
(this is the so-called QESS, quasi-exactly solvable system).
Therefore, for the spin $S=0$ we can find only the ground state wave
function and eigenvalue:
\be
\Psi_0 (\zeta) = C_1 
\exp{
    \left(
           - \frac{B}{2} \cosh{\zeta}
    \right)
    },\ 
\varepsilon_0 = 0.
\ee
Hence, we obtain that the ground-state mass of 
the quantum particle with curvature coincides with the classical one,
\be                                                         \lb{eq4.18}
\Mass_0 = \mu,
\ee
as was expected.

Further, in absence of exact wave functions for more excited 
levels one can find
the first (small) quantum correction to mass 
in the approximation of the quantum harmonic oscillator.
It is easy to see that at $B \geq 1$
the (effective) potential 
\be
V(\zeta) = 
\left(
      \frac{B}{2}
\right)^2 \text{sinh}^2 \zeta
-
\frac{B}{2} \cosh{\zeta}
\ee
has the single minimum
\[
V_{\text{min}} = - B/2 \ \ \text{at} \ \ \zeta_{\text{min}}=0.
\]

Then following the $\hbar$-expansion technique we shift the origin of 
coordinates in the point of minimum (to satisfy 
$\varepsilon = \varepsilon_0 = 0$ in absence
of quantum oscillations), and expand $V$ in the 
Taylor series to second order
near the origin thus reducing the model to the oscillator
of the unit mass, energy $\varepsilon/2$ and oscillation frequency
\[
\omega = \frac{1}{2} \sqrt{B(B-1)}.
\]
Therefore, the quantization rules yield the discrete spectrum
\be
\varepsilon =  \sqrt{B(B-1)} (n + 1/2) 
+ O (\hbar^2),\ \ n=0,\ 1,\ 2, ...,
\ee
and the first quantum correction to particle masses will be
determined by the lower energy of oscillations:
\be                                                        
\varepsilon = \frac{1}{2} \sqrt{B(B-1)} + O (\hbar^2),
\ee
that gives the algebraic equation for $\Mass$ as a function of $m$
and $\mu$.

We can easily resolve it in the approximation 
\be                                                        \lb{eq4.22}      
B \gg 1 \ \Adequa \ c/\mu^2 \to 0,
\ee
which is admissible for the major physical cases, and obtain
\be                                                         
\varepsilon = \frac{B}{2} + O (\hbar^2 c/\mu^2),
\ee
that after considering of eqs. (\ref{eq4.13}) and (\ref{eq4.18}) yields
\be
(\Mass-\mu)^2 = \frac{c \Mass}{4 \mu} + O (\hbar^2 c/\mu^2).
\ee
Then one can seek for mass in the form $\Mass=\mu+\delta$ 
($\delta \ll \mu$), and 
finally we obtain the mass of a particle with curvature (\ref{eq3.22})
with first-order quantum corrections
\be                                             \lb{eq4.25}
\Mass = \mu \pm \frac{\sqrt{c}}{2} + O(\hbar^2 c/\mu^2).
\ee
The nature of the justified choice of the root sign before the second term
is not so clear as it seems for a first look,
because there exist the two historically interfering arguments.
The first (physical) one is: 
if we apply this formalism for the one-scalar $\varphi^4$
model \cite{kpp} and compare the result with that obtained in other
ways \cite{raj}, we should suppose the sign ``$+$''.
However, the second, mathematical, counterargument is as follows: 
the known exact spectra of the operators with the QES potentials 
like (\ref{eq4.11}) are
split, as a rule by virtue of radicals, hence the signs ``$\pm$''
can approximately represent such a bifurcation and thus
should be unharmed.
If it is really so, quantum fluctuations should divide the 
classically unified particle with
curvature into several subtypes with respect to mass.

Finally, comparing the first term (\ref{eq4.25}) 
and the estimate (\ref{eq4.22}), one
can see that the obtained spectrum is 
nonperturbative and can not be derived by virtue of the Taylor
series with respect to $1/\mu$.

\subsection{Mass of quantum JT black hole}\lb{s-q-mo}

Thus, considering eqs. (\ref{eq3.35}) and (\ref{eq4.25}), the mass of 
quantum JT black hole as a soliton-dilaton boson 
in the first approximation is
\be
\Mass = M \pm m/2 + O(m^2/M^2),
\ee
therefore, the approximation (\ref{eq4.22}) has to be justified
in this case.
The problem of the obtaining of further corrections turns out to be
reduced to the mathematically standard
Sturm-Liouville problem for the Razavi potential, all 
the more so the latter is well-like on the whole axis and hence admits
only the bound states with a discrete spectrum.

Finally, it should be noted that we quantized the reduced 
theory (\ref{eq3.26}) rather than complete dilaton gravity because in 
general case the latter 
has two first-class constraints which were removed by fixed metric gauge.
Besides, unlike the previous works we quantized the theory about the
static solution rather than in the neigbourhood of trivial vacuum, 
and were interested first of all in obtaining of mass spectrum.
The question of the construction of corresponding formalism for dilaton
gravity in general case remains open because it requires 
the consistent generalization of the
field-to-particle transition procedure for 
fields in curved spacetime.

\section{Conclusion}\lb{s-c}

Let us enumerate the main items studied.
It was shown that the Jackiw-Teitelboim dilaton gravity can
be reduced to biscalar theory admitting the 
doublet consisting of instanton and dilaton components,
which can be interpreted as a massive quantum particle.
Further, considering field fluctuations in the neighborhood of 
the JT black hole it was ruled out the action for JT field
doublet as a non-minimal point particle with curvature,
thereby we generalized the procedure of
obtaining of brane actions for the multiscalar case.
From the fact, that the (1+1)-dimensional dilaton gravity
yields the effective action for JT black hole as a spatially
zero-dimensional brane (nonminimal point particle),
we can conclude that the ordinary 4D
black hole (in the case of arbitrary symmetry and field
fluctuations in a neighborhood) could be consecutively described within
frameworks of a five-dimensional field theory.

When quantizing this action as the constrained 
theory with higher derivatives, 
it was shown that the resulting \schrod equation is the
special case of that with the Razavi
potential having SU(2) dynamical symmetry group in the ground state.
Finally, we found the first quantum correction to 
mass of the Jackiw-Teitelboim black hole which could not be 
calculated by means of the perturbation theory.

\appendix
\section{Eigenvalue theorem}\lb{a-a}

{\it Theorem.}
The bound-state singular Sturm-Liouville problem
\be                                               \lb{eq-a1}
-f''(u) +  
\left(
      1 - 2\, \text{sech}^2 u
\right)f(u) - c f(u)  = 0,
\ee
\be                                               \lb{eq-a2}
f(+\infty) = f(-\infty) = O(1),
\ee 
has only the two sets of eigenfunctions and eigenvalues
\ba
&&f_0 = K_0\, \text{sech}\, u,\ c_0=0,           \nn\\
&&f_1 = K_1 \tanh{u},\ c_1=1.                \nn
\ea
where $K_i$ are arbitrary integration constants.

{\it Proof.}
Performing the change $z =  \text{cosh}^2 u$, we rewrite the 
conditions of the theorem in the form
\be                                                \lb{eq-a3}
2 z (z-1) f_{z z} + (2 z -1) f_{z} -
\left(
      \frac{\widetilde c}{2} - \frac{1}{z}
\right) f = 0,
\ee
\be                                               \lb{eq-a4}
f(1) = 0,\ \ f(+\infty) = O(1),
\ee 
where $\widetilde c = 1 - c$.
The general integral of eq. (\ref{eq-a3}) 
can be expressed in terms of the hypergeometric functions
\[
f = 
\frac{C_1}{\sqrt{z}}
\HypGF{\frac{-1-\sqrt{\widetilde c}}{2}}
      {\frac{-1+\sqrt{\widetilde c}}{2}}
      {-\frac{1}{2}}{z}
+ C_2 z
\HypGF{1-\frac{\sqrt{\widetilde c}}{2}}
      {\frac{1+\sqrt{\widetilde c}}{2}}
      {\frac{5}{2}}{z}.
\]
Using the asymptotics of the hypergeometric functions in the neighborhood
$z=1$, it is straightforward to derive that the first from the conditions
(\ref{eq-a4}) will be satisfied if we suppose
\be                                                         \lb{eq-a5}
\frac{1}{C_1} f^{(\text{reg})} = 
\frac{1}{\sqrt{z}}
\HypGF{-1-\frac{\sqrt{\widetilde c}}{2}}
      {-1+\frac{\sqrt{\widetilde c}}{2}}
      {-\frac{3}{2}}{z}
- C^{(\text{reg})} z
\HypGF{\frac{3-\sqrt{\widetilde c}}{2}}
      {\frac{3+\sqrt{\widetilde c}}{2}}
      {\frac{7}{2}}{z},
\ee
where
\[
C^{(\text{reg})} = \sqrt{\widetilde c}  
(\widetilde c - 1)
\tan{
\left(
    \frac{\pi\sqrt{\widetilde c}}{2}
\right)
    }.
\]
Further, to specify the parameters at which this function satisfies
with the second condition (\ref{eq-a4}) we should consider the 
asymptotical behavior of $f^{(\text{reg})}$ near infinity.
We have
\be                  
\frac{1}{C_1} f^{(\text{reg})} (z \to \infty) =
\frac{2\, \breve\gamma}{\pi^{3/2}}
(-1)^{1+\sqrt{\widetilde c}/2} 
\tan{
     \left( 
           \frac{\pi \sqrt{\widetilde c}}{2}
     \right)
    }
\sin{
     \left( 
           \frac{\pi \sqrt{\widetilde c}}{2}
     \right)
    }
z^{\sqrt{\widetilde c}/2}
\biggl[ 
       1+ O(1/z)
\biggr],
\ee
where
\[
\breve\gamma = \Gamma (\sqrt{\widetilde c})
\left[
      i \sqrt{\widetilde c} (\widetilde c - 1) 
      \Gamma (-1/2 - \sqrt{\widetilde c}/2)
      \Gamma (\sqrt{\widetilde c}/2)
       - 8\,
      \Gamma (1 - \sqrt{\widetilde c}/2)
      \Gamma (3/2 - \sqrt{\widetilde c}/2)
\right].
\]
From this expression it can easily be seen that 
$ f^{(\text{reg})}$ diverges at infinity everywhere except
perhaps the points:
\[
\widetilde c = (2 n)^2 = 0,\ 4,\ 16, ...,\ \ 
\text{and} \ \ \widetilde c = 1,
\]
which demand on an individual consideration.
From eq. (\ref{eq-a3}) we have 
\[
f_{\widetilde c = 0} = 
C_1 
\sqrt{1-\frac{1}{z}} 
+ 
C_2 
\left[
      i - 
      \sqrt{1-\frac{1}{z}} 
      \arcsin{\sqrt{z}}
\right],
\]
\[
f_{\widetilde c = 1} = 
\frac{C_1}{\sqrt{z}} 
+ 
C_2 
\left[
      \sqrt{1-\frac{1}{z}} -
      i \frac{\arcsin{\sqrt{z}}}{\sqrt{z}}
\right],
\]
\[
\widetilde f_{\widetilde c = 4} = 
C_1 
\sqrt{1-\frac{1}{z}} 
\left(
      2 z + 1
\right)
+ C_2 z,
\]
\[
f_{\widetilde c = 16} = 
C_1 
\sqrt{1-\frac{1}{z}} 
\left(
      24 z^2 - 8 z - 1
\right)
+ C_2 z (1-6 z/5),
\]
and so on.
By induction it is clear that at 
$\widetilde c \geq 4$
there are no $C_i$ at which $ f$ would satisfy with the requirements
(\ref{eq-a2}).

\def\AnP{Ann. Phys.}
\def\APP{Acta Phys. Polon.}
\def\CJP{Czech. J. Phys.}
\def\CMPh{Commun. Math. Phys.}
\def\CQG {Class. Quantum Grav.}
\def\IJMP  {Int. J. Mod. Phys.}
\def\JMP{J. Math. Phys.}
\def\JPh{J. Phys.}
\def\FP{Fortschr. Phys.}
\def\GRG {Gen. Relativ. Gravit.}
\def\LMPh {Lett. Math. Phys.}
\def\MPL  {Mod. Phys. Lett.}
\def\NPh  {Nucl. Phys.}
\def\PhE  {Phys.Essays}
\def\PhL  {Phys. Lett.}
\def\PhR  {Phys. Rev.}
\def\PhRL {Phys. Rev. Lett.}
\def\PhRp {Phys. Rep.}
\def\NCim {Nuovo Cimento}
\def\TMF {Teor. Mat. Fiz.}
\def\prp {report}
\def\Prp {Report}

\def\jn#1#2#3#4#5{{#1}{#2} {#3} {(#5)} {#4}}   

\def\boo#1#2#3#4#5{ #1 ({#2}, {#3}, {#4}){#5}}  



\begin{references}

\bibitem{zlo005}
Such models are very popular in general relativity because there it is
possible to take into account quantum nature through  considering of
features of some phenomenological ``macroscopical'' (quasi)matter
and to work with collective degrees of freedom using simplest symmetries. 
For some historical review and modern examples see:  
K. G. Zloshchastiev, 
\jn{\CQG}{}{16}{1737}{1999}; 
\jn{\GRG}{}{31}{1821}{1999} (updates in gr-qc/0001002);  
\jn{\IJMP}{ D}{8}{165}{1999} (updates in gr-qc/9807012).

\bibitem{jt}
R. Jackiw and C. Teitelboim, in
\boo{Quantum Theory of Gravity}{Adam Hilger}{Bristol}{1984}{}.

\bibitem{btz}
M. Ba\~nados, C. Teitelboim, and J. Zanelli,
\jn{\PhRL}{}{69}{1849}{1992}.

\bibitem{ao}
A. Ach\'ucarro and M. E. Ortiz,
\jn{\PhR}{ D}{48}{3600}{1993}. 

\bibitem{cm}
M. Cadoni and S. Mignemi,
\jn{\PhR}{ D}{51}{4319}{1995}. 

\bibitem{gk}
J. Gegenberg and G. Kunstatter,
\jn{\PhL}{ B}{413}{274}{1997}; 
\jn{\PhR}{ D}{58}{124010}{1998}. 

\bibitem{lgk}
D. Louis-Martinez, J. Gegenberg, and G. Kunstatter,
\jn{\PhL}{ B}{321}{193}{1994}.

\bibitem{dil-qua}
H. Terao,
\jn{\NPh}{ B}{395}{623}{1993};
E. Elizalde and S. Odintsov,
\jn{\NPh}{ B}{399}{581}{1993};
D. Cangemi, R. Jackiw and B. Zwiebach,
\jn{\AnP}{}{245}{408}{1996};
M. Cavagli\'a, V. de Alfaro and A. T. Filippov,
\jn{\PhL}{ B}{424}{265}{1998}.

\bibitem{kpp}
A. A. Kapustnikov, A. Pashnev, and A. Pichugin, 
\jn{\PhR}{ D}{55}{2257}{1997}. 

\bibitem{zlo006}
K. G. Zloshchastiev, 
\jn{\PhL}{ B}{450}{397}{1999}; 
\jn{\JPh}{ G}{25}{2177}{1999}.


\bibitem{ner}
R. D. Pisarski,
\jn{\PhR}{ D}{34}{670}{1986};
C. Battle, J. Gomis and N. Roman-Roy,
\jn{\JPh}{ A: Math. Gen.}{21}{2693}{1988};
M. Pa\v vsi\v c,
\jn{\PhL}{ B}{205}{231}{1988}; 
\jn{\PhL}{ B}{221}{264}{1989}; 
H. Arodz, A. Sitarz and P. Wegrzyn,
\jn{\APP}{ B}{20}{921}{1989}; 
J. Grundberg, J. Isberg, U. Lindstr\" om and H. Nordstr\" om,
\jn{\PhL}{ B}{231}{61}{1989}; 
J. Isberg, U. Lindstr\" om and H. Nordstr\" om,
\jn{\MPL}{ A}{5}{2491}{1990};
A. Dhar,
\jn{\PhL}{ B}{214}{75}{1988}; 
M. Huq, P. I. Obiakor and S. Singh,
\jn{\IJMP}{ A}{5}{4301}{1990};
J. P. Gauntlet, K. Itoh and P. K. Townsend,
\jn{\PhL}{ B}{238}{65}{1990}; 
A. Nersessian, 
\jn{\TMF}{}{117}{130}{1998}. 

\bibitem{ply}
M. S. Plyushchay, 
\jn{\MPL}{ A}{3}{1299}{1988};
\jn{\MPL}{ A}{4}{837}{1989};
\jn{\PhL}{ B}{243}{383}{1990};
\jn{\NPh}{ B}{362}{54}{1991};
\jn{\PhL}{ B}{253}{50}{1991};
\jn{\PhL}{ B}{262}{71}{1991};
\jn{\MPL}{ A}{10}{1463}{1995};
Yu. A. Kuznetsov and M. S. Plyushchay,
\jn{\NPh}{ B}{389}{181}{1993}.

\bibitem{dhot}
T. Dereli, D. H. Hartley, M. Onder, and R. W. Tucker,
\jn{\PhL}{ B}{252}{601}{1990};
V. V. Nesterenko, 
\jn{\JPh}{ A: Math. Gen.}{22}{1673}{1989};
\jn{\CQG}{}{9}{1101}{1992}.

\bibitem{raz} 
M. Razavi,
\jn{Am. \JPh}{}{48}{285}{1980}.

\bibitem{zu} 
V. V. Ulyanov and O. B. Zaslavsky,
\jn{\PhRp}{}{216}{188}{1992};
H. Konwent, P. Machnikowski, and A. Radosz,
\jn{\JPh}{ A}{28}{3757}{1995}.

\bibitem{lp}
Ye. M. Lifshitz and L. P. Pitaevskii,
\boo{Statistical Physics}{Nauka}{Moskow}{1978}{}.

\bibitem{zlop}
K. G. Zloshchastiev, in preparation. 

\bibitem{raj}
R. Rajaraman,
\boo{Solitons and Instantons}{North-Holland}{Amsterdam}{1988}{}.


\end{references}
\end{document}